\documentclass{article}
\usepackage[T1]{fontenc} % add special characters (e.g., umlaute)
\usepackage[utf8]{inputenc} % set utf-8 as default input encoding
\usepackage{ismir,amsmath,amsfonts,amsthm,cite,url}
\usepackage{graphicx}
\usepackage{booktabs}
\usepackage{color}

\newtheorem{prop}{Proposition}

\usepackage[bookmarks=false,hidelinks]{hyperref}
\title{Coupled Recurrent Models for Polyphonic Music Composition}

\multauthor
{John Thickstun$^1$ \hspace{1cm} Zaid Harchaoui$^2$ \hspace{1cm} Dean P. Foster$^3$ \hspace{1cm} Sham M. Kakade$^{1,2}$} { 
  $^1$ Allen School of Computer Science and Engineering, University of Washington, Seattle, WA, USA\\
$^2$ Department of Statistics, University of Washington, Seattle, WA, USA\\
$^3$ Amazon, NY, USA \\
{\tt\small \{thickstn,sham\}@cs.washington.edu, zaid@uw.edu, foster@amazon.com}
}
\def\authorname{John Thickstun, Zaid Harchaoui, Dean P. Foster, Sham M. Kakade}

\begin{document}

\maketitle
\begin{abstract}
This paper introduces a novel recurrent model for music composition that is tailored to the structure of polyphonic music. We propose an efficient new conditional probabilistic factorization of musical scores, viewing a score as a collection of concurrent, coupled sequences: i.e. voices. To model the conditional distributions, we borrow ideas from both convolutional and recurrent neural models; we argue that these ideas are natural for capturing music's pitch invariances, temporal structure, and polyphony. We train models for single-voice and multi-voice composition on 2,300 scores from the KernScores dataset.
\end{abstract}

\section{Introduction}\label{intro}

In this work we will think of a musical score as a sample from an unknown probability distribution. Our aim is to learn an approximation of this distribution, and to compose new scores by sampling from this approximation. For a broad survey of approaches to automatic music composition, see \cite{herremans2017functional}; for a more targeted survey of classical probabilistic approaches, see \cite{conklin2003music}. We note the success of parameterized, probabilistic generative models in domains where problem structure can be exploited by models: convolutions in image generation, or autoregressive models in language modeling. This work examines autoregressive models of scores (Section~\ref{factor}): how to evaluate these models, how to build the structure of music into parameterized models, and the effectiveness of these modeling choices.

We study the impact of structural modeling assumptions via a cross-entropy measure (Section~\ref{setup}). It is reasonable to question whether cross-entropy is a good surrogate measure for the subjective quality of sampled compositions. In theory, a sufficiently low cross-entropy indicates a good approximation of the target distribution and therefore must correspond to high-quality samples. In practice, we observe of other generative modeling tasks that learned models do achieve sufficiently low cross-entropy to produce qualitatively good samples \cite{dai2018transformer,oord2016pixel,van2016wavenet}. Studying the cross-entropy allows us to explore many models with various structural assumptions (Section~\ref{models}). Finally, we provide a qualitative evaluation of samples from our best model to demonstrate that these models have sufficiently small cross-entropy for samples to exhibit a degree of subjective quality (Section~\ref{sec:conclude}). Supplementary material including appendices, compositional samples, and code for the experiments is available online.\footnote{{\color{blue}\url{http://homes.cs.washington.edu/\~thickstn/ismir2019composition/}}}
\section{Related Works}\label{related}

In this work, we consider both single-voice models and multi-voice, polyphonic models. Early probabilistic models of music focused on single-voice, monophonic melodies. The first application of neural networks to melody composition was proposed by \cite{todd1989}. This work prompted followup \cite{mozer1994} using an alternative data representation inspired by pitch geometry ideas \cite{shepard1982}; the relative pitch and note-embedding schemes considered in the present work can be seen as a data-driven approach to capturing some of these geometric concepts. For recent work on monophonic composition, see \cite{sturm2016,jaques2017,roberts2018}.

Work on polyphonic music composition is considerably more recent. Early precurors include \cite{kohonen1989}, which considers two-voice composition, and \cite{ebciouglu1988}, which proposes an expert system to harmonize 4-voice Bach chorales. The harmonization task became popular, along with the Bach chorales dataset \cite{allan2005}. Multiple voice polyphony is directly addressed in \cite{lavrenko2003}, albeit using a simplified preprocessed encoding of scores that throws away duration information. 

Maybe the first work with a fair claim to consider polyphonic music in full generality is \cite{lewandowski2012}. This paper proposes a coarse discrete temporal factorization of musical scores (for a discussion of this raster factorization and others, see Section \ref{factor}) and examines the cross-entropy of a variety of neural models on several music datasets (including the Bach chorales). Many subsequent works on polyphonic models use the dataset, encoding, and quantitative metrics introduced in \cite{lewandowski2012}, notably \cite{vohra2015} and \cite{johnson2017}. We also note recent, impressive work on the closely related problem of modeling expressive musical performances \cite{oore2018,huang2019}.

Many recent works focus exclusively on the Bach chorales dataset \cite{liang2017,hadjeres2017,huang2017}. The works \cite{liang2017,hadjeres2017} evaluate their models using qualitative large-scale user studies. The system proposed in \cite{hadjeres2017} optimizes a pseudo-likelihood, so its quantitative losses cannot be directly compared to generative cross-entropies. The generative model proposed in \cite{liang2017} could in principle report cross entropies, but this work also focuses on a qualitative study. Quantitative cross-entropy metrics on the chorales are analyzed in \cite{huang2017}. Both \cite{hadjeres2017} and \cite{huang2017} propose non-sequential Gibbs-sampling schemes for generation, in contrast to the ancestral samplers used in \cite{liang2017} and in the present work.

\section{Factoring the Distribution over Scores}\label{factor}
Polyphonic scores consist of notes and other features of variable length that overlap each other in quasi-continuous time. Scores contain a vast heterogenous collection of information, much of which we will not attempt to model: time signatures, tempi, dynamics, etc. We will therefore give a working definition of a score that captures the pitch, rhythmic, and voicing information we plan to model. We define a score of length $T$ beats as a continuous-time, matrix-valued sequence $\textbf{x}$, where $\textbf{x}_t \in \{0,1\}^{V \times 2P}$ for each time $t \in [0,T]$. Specifically, for each voice $v \in \{1,\dots,V\}$ and each pitch $p \in \{1,\dots,P\}$ we set
\begin{align}\label{eqn:notebit}
& \textbf{x}_{t,v,p} = 1 & & \text{iff pitch $p$ is on at time $t$ in voice $v$},\\
& \textbf{x}_{t,v,P+p} = 1 & & \text{iff pitch $p$ begins at time $t$ in voice $v$}.\label{eqn:onsetbit}
\end{align}
Both ``note'' bits \eqref{eqn:notebit} and ``onset'' bits \eqref{eqn:onsetbit} are required to represent a score, expressing the distinction between a sequence of repeated notes of the same pitch and a single sustained note; see Appendix C for further discussion.

Let $q$ denote the (unknown) probability distribution over scores $\textbf{x}$. Score are high dimensional objects, of which we have limited samples (2,300 -- see Section \ref{setup}). Rather than directly model $q$, we will serialize \textbf{x}, factor $q$ according to this serialization, and model the resulting conditional distributions $q(\cdot|\text{history})$. There are many possible ways to factor $q$; in the remainder of this section we review the popular raster factorization, and propose a new sequential factorization based on voices.

\textbf{Raster score factorization.} Many previous works factor a score via rasterization. If we sample a score $\textbf{x}$ at constant intervals $\Delta$ and impose an order on parts and notes, we can factor the distribution $q$ over scores as $q(\textbf{x}) = $
\begin{equation}\label{eqn:raster}
 \prod_{k=1}^{T/\Delta} \prod_{v=1}^V\prod_{p=1}^{2P} q(\textbf{x}_{k\Delta,v,p}|\textbf{x}_{1:k\Delta},\textbf{x}_{k\Delta,1:v},\textbf{x}_{k\Delta,v,1:p}).
\end{equation}
Throughout this work, a slice \texttt{$1$:$i$} is inclusive of the first index $1$ but does not include the final index $i$. 

This factorization generates music in sequential $\Delta$-slices of time. Some prior works directly model the (high-dimensional) distribution $\textbf{x}_{k\Delta}$; this approach was pioneered by \cite{lewandowski2012}, using NADE to model the conditional distributions $q(\textbf{x}_{k\Delta}|\textbf{x}_{1:k\Delta})$. Others impose further order on notes (and voicings, if they choose to model them) and factor the distribution into binary conditionals as in \eqref{eqn:raster}. Notes are typically ordered based on pitch, either low-to-high \cite{hadjeres2017} or high-to-low \cite{liang2017}.

\textbf{Sequential voice factorization.} Putting full scores aside for now, consider factoring a single voice $v$, i.e. a slice $\textbf{x}_{1:T,v,1:2P}$ of a score. By definition, a KernScores voice is homophonic in the sense that its rhythms proceed in lock-step: a voice consists of a sequence of notes, chords, or rests, and no notes are sustained across a change point.\footnote{For polyphonic instruments like the piano, we must adopt a more refined definition of a voice than ``notes assigned to a particular instrument;'' see Appendix B for details.} Instead of generating raster time slices, suppose we run-length encode the durations between change points in $v$. We denote these change points by $c^v_0,\dots,c^v_{L_v}$ where $L_v$ is the number of change points in voice $v$. Let $D$ be the number of unique distance between change points, and define a run-length encoded voice $\textbf{r} \in \left(\{0,1\}^D \oplus \{0,1\}^N\right)^{L_v}$. At each index $k \in \{1,\dots,L_v\}$, $\textbf{r}_k = (\textbf{r}_{k,0},\textbf{r}_{k,1})$ with $\textbf{r}_{k,0} \in \{0,1\}^D$ and $\textbf{r}_{k,1} \in \{0,1\}^N$ such that
\begin{align*}
& \textbf{r}_{k,0} = \textbf{1}_{d_k} & & \text{where $d_k = \frac{c^v_{k+1} - c^v_{k}}{\Delta} \in \mathbb{N}$},\\
& \textbf{r}_{k,1,p} = 1 & & \text{iff pitch $p$ begins at time $c_k^v$ in voice $v$}.
\end{align*}
The durations $d_k$ correspond to note-values (quarter-note, eighth-note, dotted-half, etc.). We proceed to factor the voice sequentially as $p(\textbf{r}) = $
\begin{equation}\label{eqn:partfactor}
 \prod_{k=1}^{L_v} q(\textbf{r}_{k,0}|\textbf{r}_{1:k}) \prod_{p=1}^P q(\textbf{r}_{k,1,p}|\textbf{r}_{1:k},\textbf{r}_{k,0},\textbf{r}_{k,1,1:p}).
\end{equation}

\textbf{Sequential score factorization.} We now consider a sequential factorization that interlaces predictions in the score's constituent voices. The idea is to predict voices sequentially as we did in the previous section, but we must now choose the order across voices in which we make predictions. The rule we choose is to make a prediction in the voice that has advanced least far in time, breaking ties by the arbitrary numerical order assigned to voices (ties happen quite frequently: for example, at the beginning of a score when all parts have advanced 0 beats). This ensures that all voices are generated in near lock-step; generation in any particular voice never advances more than one note-value ahead of any other voice.

Mathematically, we can describe this factorization as follows. First, we impose a total order on change points $c^v_k$ across voices by the rule $c^v_k < c^u_{k'}$ for all $v,u$ if $k < k'$ and $c^v_k < c^u_k$ if $v < u$. Define $L \equiv \sum_{v=1}^V L_v$. For index $i \in \{1,\dots,L\}$ let $\alpha_i$ and $\beta_i$ denote the index and voice of the corresponding change point according the the total ordering on change points. We define a sequentially encoded score $\textbf{s} \in (\{0,1\}^D \oplus \{0,1\}^N)^L$ by
\begin{align*}
& \textbf{s}_{k,0} = \textbf{1}_{d_k} & & \text{where $d_k = \frac{c_{\alpha_k+1}^{\beta_k} - c_{\alpha_k}^{\beta_k}}{\Delta} \in \mathbb{N}$},\\
& \textbf{s}_{k,1,p} = 1 & & \text{iff pitch $p$ begins in voice $\beta_k$ at time $c_{\alpha_k}^{\beta_k}$}.
\end{align*}
And we factor the distribution sequentially by $q(\textbf{s}) = $
\begin{equation}\label{eqn:scorefactor}
 \prod_{k=1}^L q(\textbf{s}_{k,0}|\textbf{s}_{1:k}) \prod_{p=1}^P q(\textbf{s}_{k,1,p}|\textbf{s}_{1:k},\textbf{s}_{k,0},\textbf{s}_{k,1,1:p}).
\end{equation}

This factorization produces a ragged frontier of generation, where generation in a particular part advances no further than one note-value ahead of the other parts at any point in the generative process. This stands in contrast to the raster factorization, for which generation advances with a smooth frontier, one $\Delta$-slice of time after another.

\textbf{Other Factorizations.} The factorizations presented above are not comprehensive. Another alternative is a direct run-length encoding of scores, discussed in Appendix D. We could also consider alternative total orderings of change points, generating a measure or entire voice at a time in each voice. The choice of factorization has broad implications for both the computational efficiency and the parameterization of a generative model of scores; the importance of this choice in the construction of a model should not be overlooked.

\section{Dataset and Evaluation}\label{setup}

\textbf{Dataset}. The models presented in this paper are trained on KernScores data \cite{sapp2005}, a collection of early modern, classical, and romantic era digital scores assembled by musicologists and researchers associated with Stanford's CCARH.\footnote{\url{http://kern.ccarh.org/}} The dataset consists of over 2,300 scores containing approximately 2.8 million note labels. Tables~\ref{composers}~and~\ref{notes} give a sense of the contents of the dataset.

{\tiny
\begin{table*}[h]
  \setlength{\tabcolsep}{4pt}
  \centering
  \begin{tabular}{rrrrrrrrr}
    \toprule
    \cmidrule{1-9}
     Bach & Beethoven & Chopin & Scarlatti & Early & Joplin & Mozart & Hummel & Haydn  \\
     \midrule    
     191,374 & 476,989 & 57,096 & 58,222 & 1,325,660 & 43,707 & 269,513 & 3,389 & 392,998 \\
    \bottomrule
  \end{tabular}
  \caption{Notes in the KernScores dataset, partitioned by composer. The ``Early'' collection consists of Renaissance vocal music; a plurality of the Early music is composed by Josquin. }\label{composers}
\end{table*}}

We contrast this dataset's Humdrum encoding with the MIDI encoded datasets used by most works discussed in this paper.\footnote{A notable exception is \cite{lavrenko2003}, which uses data derived from the KernScores collection considered here.} MIDI was designed as a protocol for communicating digital performances, rather than digital scores. This is exemplified by the MAPS \cite{emiya2010} and MAESTRO \cite{hawthorne2018} datasets, which consist of symbolic MIDI data aligned to expressive performances. While this data is symbolic, it cannot be interpreted as scores because it is unaligned to a grid of beats and does not encode note-values (quarter-note, eighth-note, etc). Some MIDI datasets are aligned to a grid of beats, for example MusicNet \cite{thickstun2017}. But heuristics are still necessary to interpret this data as visual scores. For example, many MIDI files encode ``staccatto'' articulations by shortening the length of notes, thwarting simple rules that identify note-values based on length.

{\tiny
\begin{table}[h]
  \centering
  \begin{tabular}{rrr}
    \toprule
    \cmidrule{1-3}
     Vocal & String Quartet & Piano \\
     \midrule   
     1,412,552 & 820,152 & 586,244 \\
     \bottomrule
  \end{tabular}
  \caption{Notes in the KernScores dataset, partitioned by ensemble type.}\label{notes}
\end{table}}

\textbf{Evaluation}. Let $\hat q$ be an estimate of the unknown probability distribution over scores $q$. We want to measure the quality of $\hat q$ by its cross-entropy to $q$. Because the entropy of a score grows with its length $T$, we will consider a cross-entropy rate. By convention, we measure time in units of beats, so the cross-entropy rate has units of bits per beat.

Defining cross-entropy for a continuous-time process generally requires some care. But for music, defining the cross-entropy on an appropriate discretization will suffice. Musical notes begin and end at rational fractions of the beat, and therefore we can find a common denominator $d$ of all change points in the support of the distribution $q$ (for our dataset $d=48$). For a score of length $T$ beats, we partition the interval $[0,T]$ into constant subintervals of length $\Delta \equiv 1/d$ and define a rate-adjusted, discretized cross-entropy
\begin{equation}\label{def:crossent}
H_\mathcal{P}(q||\hat q) \equiv  \underset{\textbf{x} \sim q}{\mathbb{E}}\left[-\frac{1}{T\Delta}\log \hat q(\textbf{x}_0,\textbf{x}_\Delta,\textbf{x}_{2\Delta},\dots,\textbf{x}_{T})\right].
\end{equation}
Proposition 1 in Appendix F shows that we can think of $\Delta$ as the resolution of the score process, in the sense that any further refinement of the discretization $d$ yields no further contributions to the cross entropy. 

Definition \ref{def:crossent} is independent of any choice about how we factor $q$: it is a cross entropy measure of the joint distribution over a full score. As we discussed in Section \ref{factor}, there are many ways to factor a generative model of scores. These choices lend themselves to different natural cross-entropies, each with their own units. By measuring in units of bits per beat at the process resolution $\Delta$ as defined by Definition \ref{def:crossent}, we can compare results under different factorizations.

\textbf{Computational cost}. Raster models are expensive to train and evaluate on rhythmically diverse music. A raster model must be discretized at the process resolution $\Delta$ to generate a score with precise rhythmic detail. The process resolution $\Delta$ of a corpus containing both triplets and sixty-fourth notes is $\Delta = 3 \times 16 = 48$ positions per beat. Corpora with quintuplet patterns require a further factor of $5$, resulting in $\Delta = 240$. To generate a score from a raster factorization requires $\Delta$ predictions per beat; to ease the computational burden of prediction, when the raster approach is taken scores are typically discretizing at either 1 or 2 positions per beat \cite{lewandowski2012}. Unfortunately, this discretization well above the process resolution leads to dramatic rhythmically simplification of scores (see Appendix C).

In contrast, a sequential factorization such as \eqref{eqn:partfactor} or \eqref{eqn:scorefactor} requires predictions proportional to the average number of notes per beat, while maintaining the rhythmic detail of a score. The KernScores single-voice corpus averages $\approx 1.25$ notes per beat, requiring $1.25$ predictions per beat for sequential factorization versus $\Delta$ predictions per beat for raster factorization. The KernScores multi-voice corpus averages $\approx 5$ notes per beat, requiring $5$ predictions per beat for sequential factorization, an order of magnitude less than the $\Delta \approx 50$ predictions per beat required for raster prediction.
\section{Models and Weight-Sharing}\label{models}

Modeling voices allows us to think of the polyphonic composition problem as a collection of correlated single-voice composition problems. Learning the marginal distribution over a single voice $v$ is similar in spirit to classical monophonic tasks. Learning the distribution over KernScores voices generalizes this classical task to allow for chords: formally, a monophonic sequence would require the vector $\textbf{r}_{k,1} \in \{0,1\}^N$ described in Section \ref{factor} to be one-hot, whereas our our dataset includes voices where this vector is multi-hot, expressing intervals and chords (e.g. chords in the left hand of a piano, or double-stops for a violin).

We will explore two modeling tasks. First we consider a single-voice prediction task: learn the marginal distribution over a voice $v$, estimating the conditionals that appear in the factorization \eqref{eqn:partfactor}. Results on this task are summarized in Table \ref{homophonicresults}. Second we consider a multi-voice prediction task: learn the joint distribution over scores, estimating the conditionals that appear in the factorization \eqref{eqn:scorefactor}. Results on this task are summarized in Table \ref{polyphonicresults}. 

\iffalse
\begin{figure}[h]
\begin{minipage}{0.5\textwidth}
\includegraphics[width=1.0\linewidth,scale=1.0]{haydn.png}
\end{minipage}
\begin{minipage}{1.0\textwidth}
\setlength\tabcolsep{12pt}
\begin{tabular}{ l | l | l | l }
1.50 : 70 & 1.00 : 58 & 1.00 : 62 & 1.00 : 58 \\
*     & 1.00 :      & 1.00 :      & 1.00 : 55 \\
0.25 : 72 &   *      & *         & * \\     
0.25 : 74 &   *      & *         & * \\
0.25 : 75 & 1.00      & 0.50 :    & 1.00 : 51 \\
0.25 : 77 &   *       & *         & * \\
0.25 : 79 &    *      & 0.25 : 75   & * \\
0.25 : 81 &     *     & 0.25 : 77    & *  \\

1.00 : 82 & 3.00 : & 1.00 : 79 & 1.50 : 46 \\
... & ... & ... & ...
%1.0 : 81 & * : & 1.0 : 77 & * : 46\\
%* : 81 & * : & * : 77 & 0.25 : 48\\       
%* : 81 & * : & * : 77 & 0.25 : 50\\      
%1.0 : 79 & * : & 1.0 : 75 & 0.25 : 51 \\
%* : 79 & * : & * : 75 & 0.25 : 53 \\
%* : 79 & * : & * : 75 & 0.25 : 55 \\  
%* : 79 & * : & * : 75 & 0.25 : 57 \\
\end{tabular}
\end{minipage}
\caption{Top: a partial score for Haydn's string quartet Opus 55 No. 3, first movement, from measure 16. Bottom: a description of the history tensor $\textbf{s}$ for the Haydn quartet above. A frame of time is indicated by a row. Each event in a voice is denoted by a pair of duration and note(s), separated by a colon. Durations are denominated in beats. An asterisk indicated continuation of the previous note(s). }\label{fig:polyencoding}
\end{figure}
\fi

\subsection{Representation}\label{sec:repr}

Like our choice of factorization, we are faced with many options for encoding the history of a score for prediction. Some of the same computational and modeling considerations apply to both the choice of a factorization and the choice of a history encoding, but these are not inherently connected decisions. For the single-voice task, we use the encoding \textbf{r} introduced to define the sequential voice factorization in Section \ref{factor}.

For the polyphonic task, we also encode history using a run-length encoding. Let $c_1,\dots,c_K$ denote change points in the full score \textbf{x}, let $d^v_j \equiv (c^v_{j+1} - c^v_{j})/\Delta \in \mathbb{N}$, and define a sequence $\textbf{e} \in (\{0,1\}^{D+1} \oplus \{0,1\}^P)^{K \times V}$ where
\begin{align*}
& \textbf{e}_{k,v,0,0:D} = \textbf{1}_{d_j^p} & & \text{iff $c_k = c^v_j$ for some $c^v_j$ in voice $v$},\\
& \textbf{e}_{k,v,0,D} = 1 & & \text{iff $c_k$ is not a change point in voice $v$},\\
& \textbf{e}_{k,v,1,p} = 1 & & \text{iff pitch $p$ begins in voice $v$ at time $c_k$}.
\end{align*}

This is not the fully serialized encoding \textbf{s} used to define a score factorization (for discussion of a fully sequential representation, see \cite{oore2018}). At each time step $k$ for which any voice exhibits a change point, we make an entry in \textbf{e} for every voice; we refer to $\textbf{e}_k$ as a frame. This requires us to augment our alphabet of duration symbols $D$ with a special continuation symbol that indicates no change (comparable to the onset bits in the encoding \textbf{x}). An advantage of this representation over sequential or raster representations is that more history can be encoded with shorter sequences.

For a fixed voice $v$, let $\tilde{\textbf{r}} \equiv \textbf{e}_{:,v}$ be a single-voice slice of the score history. Observe that $\tilde{\textbf{r}} \neq \textbf{r}$, where \textbf{r} is the run-length encoding used for the single-voice task. The slices $\tilde{\textbf{r}}$ are spaced out with aforementioned continuation symbols where there are change points in other voices. With the single-voice encoding \textbf{r}, simple linear filters can be learned that are sensitive to particular rhythmic sequences: e.g. groups of four eighth notes, or three triplet-quarter notes. This is not the case for $\tilde{\textbf{r}}$; rhythmic patterns can be somewhat-arbitrarily broken up by continuation symbols. 

These observations might lead us to consider raster encodings for multi-voice history, which restore the effectiveness of simple linear filters at the cost of increasing the dimensionality of the history encoding. We find that recurrent networks for the single-voice task are unhampered when retrained on $\tilde{\textbf{r}}$: compare experiments 21 and 22 in Table \ref{homophonicresults}. Performance falls slightly when learning on $\tilde{\textbf{r}}$, but this is to be expected because history interspersed with continuations is effectively a shorter-length history.

For both the single-voice and multi-voice tasks, we truncate the history at a fixed number of frames prior to the prediction time. We explore several history lengths in the experiments and observe diminishing improvement in quantitative results for windows beyond the range of 10-20 frames of $\textbf{e}$: see experiment group (1,2,6,7) in Table \ref{polyphonicresults}.

\subsection{Single-voice models}\label{sec:monomodels}

Using factorization \eqref{eqn:partfactor}, we explore fully connected, convolutional, and recurrent models for learning the conditional distributions $q(\textbf{r}_{k,0}|\textbf{r}_{1:k})$ over note-values and $q(\textbf{r}_{k,1,n}|\textbf{r}_{1:k},\textbf{r}_{k,0},\textbf{r}_{k,1,1:p})$ over pitches. We build separate models to estimate $\textbf{r}_{k,0}$ and $\textbf{r}_{k,1,p}$, with respective losses Loss$_t$ and Loss$_n$. In the remainder of this section, we consider opportunities to exploit structure in music by sharing weights in our models. Quantitative results for single-voice models are summarized in Table \ref{homophonicresults}, with additional details available in Appendix A.

\textbf{Autoregressive modeling.} To build a generative model over conditionally stationary sequential data, it often makes sense to make the autoregressive assumption $q(\textbf{r}_k|\textbf{r}_{1:k}) = q(\textbf{r}_{k'}|\textbf{r}_{1:k'})$ for all $k,k' \in \mathbb{N}$. We can then learn a single conditional approximation $\hat q(\textbf{r}_k|\textbf{r}_{1:k})$ and share model parameters across all time translations.

Scores are not quite conditionally stationary; the distribution of notes and rhythms varies substantially depending on the sub-position within a beat. To address this, we follow the lead of \cite{johnson2017} and \cite{hadjeres2017} and augment our history tensor with a one-hot location feature vector $\ell$ that indicates the subdivision of the beat for which we are presently making predictions.\footnote{Location can always be computed from a full history. But we truncate the history, so this information is lost unless it is explicitly reintroduced.} Compare the loss of duration models (Loss$_t$) with and without these features in experiment pairs (3,4), (6,7), (10,11), (12,13), and (15,16).

\begin{figure*}[h]
\includegraphics[width=1.0\linewidth,scale=1.0]{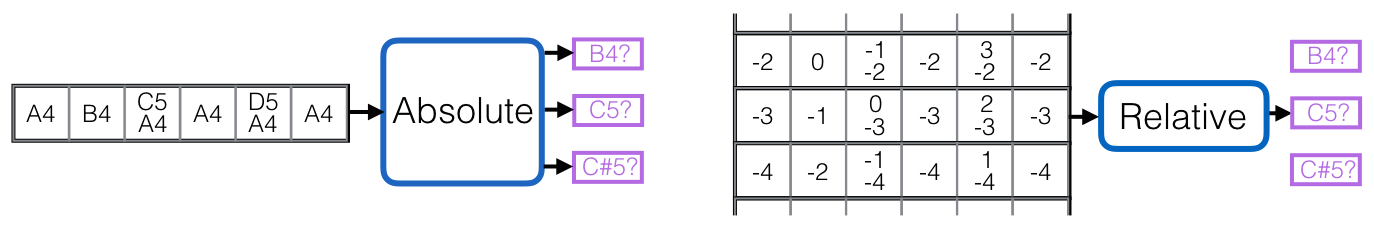}
\caption{Left: an absolute pitch predictor learns individual classifiers for each pitch-class. Right: a relative pitch predictor learns a single classifier and translates the data along the frequency axis to center it around the pitch to be predicted. Whereas the absolute predictor decides whether C5 is on given the previous note was A4, the relative predictor decides whether the note under consideration is on given the previous note was 3 steps below it.}\label{relpitch}
\end{figure*}

\textbf{Relative pitch.} We can perform a similar weight-sharing scheme with pitches as we did with time. Instead of building an individual predictor for each pitch conditioned on the notes in the history tensor, we adopt an idea proposed in \cite{johnson2017}: build a single predictor that conditions on a shifted version of the history tensor centered around the note we want to predict. Convolving this predictor over the pitch axis of the history tensor lets us make a prediction at each note location, as visualized by Figure \ref{relpitch}.

As with time, the distribution over notes is not quite conditionally stationary. For example, a truly relative predictor would generate notes uniformly across the note-class axis, whereas the actual distribution of notes concentrates around middle C. Therefore we augment our history tensor with a one-hot feature vector $\textbf{1}_p$ that indicates the pitch $p$ for which we are making a prediction. This allows us to take full advantage of all available information when making a prediction, while borrowing strength from shared harmonic patterns in different keys or octaves. We compare absolute pitch-indexed classifiers ($\textbf{lin}_p$) to a single, relative pitch classifier ($\textbf{lin}$) in Table \ref{homophonicresults}: compare the loss of pitch models (Loss$_p$) in experiment groups (2,3,4), (5,6,7), (8,9,10), (11,12,13), and (15,16).

Relative pitch models serve a similar purpose to key-signature normalization \cite{liang2017} or data augmentation via transposition \cite{hadjeres2017}. Building this invariance into the model offers an alternative approach, avoiding data preprocessing or the introduction of hyper-parameters. We find that training with transpositions in the range $\pm 5$ semi-tones yields no performance increase for relative pitch models.

\textbf{Pitch embeddings.} Borrowing the concept of a word embedding from natural language processing, we consider learned embeddings \textbf{c} of the pitch vectors $\textbf{r}_{k,1}$ ($\textbf{e}_{k,v,1}$ for the multi-voice models). For recurrent models, we do not see performance benefits to learning these embeddings: compare experiments 20 and 21 in Table \ref{homophonicresults}. However, we do find that we can learn compact embeddings (16 dimensions for the experiments presented in this paper) without sacrificing performance, and using these embeddings reduces computational cost. We also find that using a 12 dimensional fixed embedding of pitches \textbf{f}, in which we quotient each pitch class by octave, reduces overfitting for the rhythmic model while preserving predictive accuracy.

{\tiny
\begin{table}[h!]
  \centering
  \setlength\tabcolsep{2pt}
  \begin{tabular}{rclccccr}
    \toprule
    \cmidrule{1-8}
    \# & History & Arch & Loc? & Relative? & Pitch? & Embed? & Loss  \\
    \midrule
    1 & $\textbf{r}_{(1)}$ & bias & no & no & no & no & 10.07  \\
    2 & $\textbf{r}_{(1)}$ & linear & no & no & no & no & 8.05  \\
    3 & $\textbf{r}_{(1)}$ & linear & no & yes & no & no & 6.29  \\
    4 & $\textbf{r}_{(1)}$ & linear & yes & yes & yes & no & 6.12  \\
    5 & $\textbf{r}_{(1)}$ & fc &  no & no & no & no & 5.92  \\
    6 & $\textbf{r}_{(1)}$ & fc & no & yes & no & no & 6.05  \\
    7 & $\textbf{r}_{(1)}$ & fc & yes & yes & yes & no & 5.70  \\        
    \midrule
    \midrule
    8 & $\textbf{r}_{(5)}$ & linear & no & no & no & no & 7.91  \\
    9 & $\textbf{r}_{(5)}$ & linear & no & yes  & no & no & 5.76  \\
    10 & $\textbf{r}_{(5)}$ & linear & yes & yes & yes & no & 5.63  \\
    11 & $\textbf{r}_{(5)}$ & fc & no & no & no & no & 4.90  \\
    12 & $\textbf{r}_{(5)}$ & fc & no & yes & no & no & 4.80  \\
    13 & $\textbf{r}_{(5)}$ & fc & yes & yes & yes & no & 4.69  \\     
    14 & $\textbf{r}_{(5)}$ & fc & yes & yes & yes & yes & 4.63  \\ 
    \midrule
    \midrule
    15 & $\textbf{r}_{(10)}$ & linear & no & yes & no & no & 7.88 \\ 
    16 & $\textbf{r}_{(10)}$ & linear & yes & yes & yes & no & 5.53 \\ 
    17 & $\textbf{r}_{(10)}$ & fc & yes & yes & yes & yes & 4.55 \\ 
    19 & $\textbf{r}_{(10)}$ & cnn & yes & yes & yes & yes & 4.42 \\ 
    20 & $\textbf{r}_{(10)}$ & rnn & yes & yes & yes & no & 4.37 \\
    21 & $\textbf{r}_{(10)}$ & rnn & yes & yes & yes & yes & 4.36 \\
    \midrule
    \midrule
    22 & $\tilde{\textbf{r}}_{(10)}$ & rnn & yes & yes & yes & yes & 4.52 \\
    \bottomrule
  \end{tabular}
  \caption{Single-voice results. We define $\textbf{r}_{(m)} \equiv \textbf{r}_{k-m:k}$ (a truncated history of length $m$); $\tilde{\textbf{r}}_{(m)}$ is defined likewise, based on the alternate encoding $\tilde{\textbf{r}}$ discussed in Section \ref{sec:repr}, Representation. The Relative flag indicates the use of a relative-pitch classifier, and the Loc, Pitch, and Embed flags indicate the use of location features, pitch features, and pitch embeddings, discussed in Section \ref{sec:monomodels}. For additional details of these experiments, see Appendix A. } 
 \label{homophonicresults}
 \end{table}
}

\subsection{Multi-voice models}\label{sec:polymodels}

\begin{figure*}
  \centering
  \includegraphics[width=0.8\textwidth]{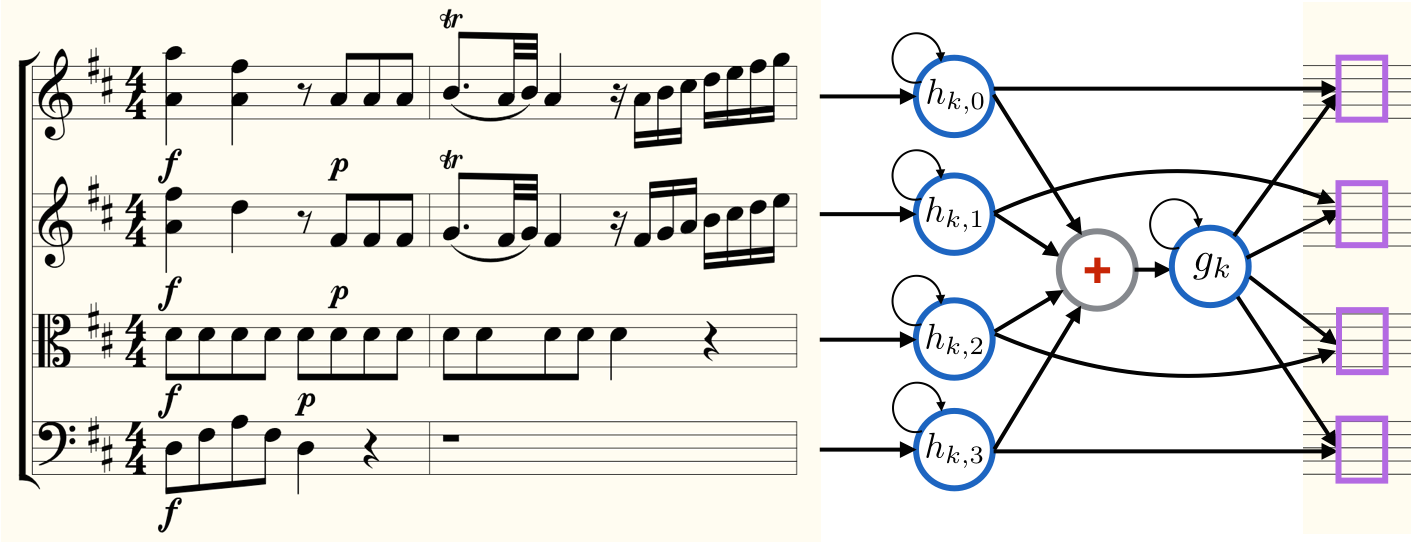}
    \caption{Coupled state estimation of Mozart's string quartet number 2 in D Major, K155, movement 1, from measure 1, rendered by the Verovio Humdrum Viewer. A recurrent network models the state $h_{k,v}$ of each voice $v$ at step $k$, based on the previous state $h_{{k-1},v}$ and the current content of the voice. Another recurrent network models of the global state $g_k$ of the score at step $k$ based on the previous global state $g_{k-1}$ and a sum of the current states of each voice. Subsequent notes (purple) in each voice are predicted using features of the global state and the state of the relevant voice. See Equations \ref{hierarchical} for a precise mathematical description of this model.}\label{concurrent_rnn}
\end{figure*}

Using the factorization \eqref{eqn:scorefactor}, we now explore ways to capture correlations between the voices and model the conditional distributions $q(\textbf{s}_{k,0}|\textbf{s}_{1:k})$ over note-values and $q(\textbf{s}_{k,1,p}|\textbf{s}_{1:k},\textbf{s}_{k,0},\textbf{s}_{k,1,1:p})$ over notes. We build separate models to estimate $\textbf{r}_{k,0}$ and $\textbf{r}_{k,1,p}$, with losses Loss$_t$ and Loss$_p$ in Table \ref{polyphonicresults} respectively. The same structural observations that we made about scores for the single-voice models apply to multi-voice modeling; all multi-voice models considered in this paper use the three weight-sharing schemes considered for single-voice models. We explore an additional weight-sharing opportunity below for the multi-voice task: voice decomposition.

The effectiveness of recurrent models for the single-voice modeling task, and the representational considerations in Section \ref{sec:repr}, motivate us to consider extensions of the recurrent architecture to capture structure in the multi-voice setting. One natural extension of the standard recurrent neural network to model multiple, concurrent voices is a hierarchical architecture, illustrated in Figure~\ref{concurrent_rnn}:
\begin{align}\label{hierarchical}
\begin{split}
& h_{k,v}(\textbf{e}) \equiv \textbf{a}\left(W_{v}^\top h_{k-1,v}(\textbf{e}) + W_e^\top \textbf{c}\left(\textbf{e}_{k,v} \right)\right),\\
& g_k(\textbf{e}) \equiv \textbf{a}\left(W_g^\top g_{k-1}(\textbf{e}) + W_{hv}^\top \sum_u h_{k,u}(\textbf{e}) \right).
\end{split}
\end{align}

The first equation is a standard recurrent network that builds a state estimate $h_{k,v}$ of a voice $v$ at time index $k$ based on transition weights $W_v$, an input embedding \textbf{c}, input weights $W_e$, and non-linear activation \textbf{a} (we use a ReLU activation). We integrate the state of each voice (weights $W_{hv}$) into a global state $g_k$ given the previous global state $g_{k-1}$ (weights $W_g$). Because voice order is arbitrary in our dataset, we sum (i.e. pool) over their states before feeding them into the global network. At each time $k$, we use the learned state of each voice together with the global state to make a note-value prediction: $\hat{\textbf{s}}_{k,0} = \textbf{lin}(h_{k,\beta_k}(\textbf{e}),g_{k}(\textbf{e}))$, where $\textbf{lin}$ is a log-linear classifier. We make pitch predictions $\textbf{s}_{k,1,p} \in \{0,1\}$ using the same architecture. We learn a single, relative-pitch classifier for $\textbf{s}_{k,1,p} \in \{0,1\}$ in all multi-voice experiments (section \ref{sec:monomodels}, Relative pitch). We do not share weights between the note-value and pitch models.

An alternate extension of a recurrent voice model to scores directly integrates the state of the other voices' states into each individual voice's state, resulting in a distributed state architecture $h_{k,v}(\textbf{e}) = $
\begin{equation}\label{distributed}
\text{a}\left(W_v^\top h_{k-1,v}(\textbf{e}) + W_x^\top \textbf{c}\left(\textbf{e}_{k,v}\right) + W_{hv}^T\sum_u h_{k,u}(\textbf{e}) \right).
\end{equation}
At each time $k$, for each voice $v$, we use the learned state of voice $v$ to make a note-value prediction $\hat{\textbf{s}}_{k,0} = \textbf{lin}(h_{k,\beta_k}(\textbf{e}))$, where $\textbf{lin}$ is a log-linear classifier. We make predictions for $\textbf{s}_{k,1,n} \in \{0,1\}$ using the same architecture and we do not share weights between the note-value and pitch models.

We find that the distributed architecture underperforms the hierarchical architecture (see Table \ref{polyphonicresults}; experiments 2 and 3) although this comparison is not conclusive. For the hierarchical model, we can consider whether the global state representation is as sensitive to history-length as the voices. Could we make successful predictions using only the final state of each voice, rather than coupling the states at each step? Experiment group (4,5,6) in Table \ref{polyphonicresults} suggests that this is not the case: we observe significant gains by integrating voice information at each time step.

Extending a loose analogy between recurrent neural networks and hidden markov models, the coupled recurrent models considered in this section could be compared to factorial hidden markov models \cite{ghahramani1996}. A crucial difference is that the distributed latent state of a coupled recurrent model is determined by the distributed input structure of a score, whereas the distributed structure of a factorial hmm only appears in the latent state.

\textbf{Voice decomposition.} Decomposing a score into multiple voices presents us with an opportunity to share weights between voice models by learning a single set of weights $W_v$ in equation (\ref{hierarchical}), rather than learning unique voice-indexed weights $W_{v_i}$ for each voice $v_i$. Indeed, because voice indices are arbitrary, the weights $W_{v_i}$ will converge to the same values for all $i$; sharing a single set of weights $W_v$ accelerates learning by enforcing this property. All score models presented in Table \ref{polyphonicresults} share these weights.

{\small
\begin{table}[h]
  \centering
  \setlength\tabcolsep{4pt}
  \begin{tabular}{rllrrr}
    \toprule
    \cmidrule{1-6}
    \# & History & Architecture & Loss & $\text{Loss}_t$ & $\text{Loss}_n$  \\
     & (voice/global) & & (total) & (time) & (notes) \\
    \midrule
    1 & 3 / 3 & hierarchical & 14.05 & 5.65 & 8.40 \\ 
    2 & 5 / 5 & hierarchical & 13.40 & 5.35 & 8.04 \\ 
    3 & 5 & distributed & 13.82 & 5.41 & 8.41 \\
    4 & 10 / 1 & hierarchical & 13.20 & 5.22 & 7.98 \\ 
    5 & 10 / 5 & hierarchical & 12.94 & 5.13 & 7.81 \\
    6 & 10 / 10 & hierarchical & 12.87 & 5.12 & 7.75 \\ 
    7 & 20 / 20 & hierarchical & 12.78 & 5.01 & 7.76 \\ 
    \midrule
    \midrule
    8 & 10 & independent & 18.63 & 6.56 & 12.08 \\ 
    \bottomrule
  \end{tabular}
  \caption{Multi-voice results. The ``hierarchical'' architecture is defined by equations \eqref{hierarchical}. Voice and global history refer to the number of time steps used to construct the states $h_{k,v}$ and $g_k$ respectively. Experiment 8 is a baseline where the voice models are completely decoupled (equivalent to single-voice Experiment 22 in Table 5; the average number of voices per score is $4.12$). Results are reported on non-piano test set data (see Appendix B for discussion of piano data). For additional experiments and ablations, see Appendix A.}
  \label{polyphonicresults}
\end{table}}

\section{Conclusion}\label{sec:conclude}

To gain insight into the quality of samples from our models, we recruited twenty study participants to listen to a variety of audio clips, each synthesized from either a real composition or from sampled output of Experiment 6 in Table \ref{polyphonicresults}. For each clip, participants were asked to judge whether the clip was written by a computer or by a human composer, following a procedure comparable to \cite{pearce2001towards}. The clips varied in length, from 10 frames of a sample \textbf{e} (2-4 seconds; the length of history conditioned on by the model) to 50 frames (10-20 seconds). Participants become more confident in their judgements of the longer clips, but even among the longest clips (around 20 seconds) participants often identified an artificial clip as a human composition. Results are presented in Table~\ref{qualitativeresults}; see Appendix E for further study details.

{\tiny
\begin{table}[h]
  \centering
  \begin{tabular}{lrrrrr}
    \toprule
    \cmidrule{1-6}
     Clip Length  &10 & 20 & 30 & 40 & 50 \\
     \midrule    
     Average & 5.3 & 5.7 & 6.6 & 6.7 & 6.8 \\
    \bottomrule
  \end{tabular}
  \caption{Qualitative evaluation of the 10-frame hierarchical model: Experiment 6 in Table \ref{polyphonicresults}. Twenty participant were asked to judge 50 audio clips each, with lengths varying from 10 to 50 frames. The scores indicate participants' average correct discriminations out of 10: 5.0 would indicate random guessing; 10.0 would indicate perfect discrimination. }
  \label{qualitativeresults}
\end{table}}

These results superficially suggest that we have done well in modeling the short-term structure of the dataset (we make no claims to have captured long-term structure; indeed, the truncated history input to our models precludes this). But it is not clear that humans are good--or should be good--at recognizing plausible local structures in music without context. See \cite{pearce2007evaluating,jordanous2012standardised} for criticism of musical Turing tests like the one presented here. It is also unclear how to use such studies to make fine-grained comparisons between models (as we have done quantitatively throughout this paper). It is not even clear how to prompt a user to discriminate between such models. Therefore we re-emphasize the interpretation of this qualitative evaluation, proposed in Section \ref{intro}, as a perceptual grounding of the quantitative evaluation considered throughout this work.

\section{Acknowledgements}

We thank Lydia Hamessley and Sreeram Kannan for sharing valuable insights. This work was supported by NSF Grants DGE-1256082, CCF-1740551, the Washington Research Foundation for innovation in Data-intensive Discovery, and the CIFAR program ``Learning in Machines and Brains.'' We also thank NVIDIA for their donation of a GPU.

% For bibtex users:
\bibliography{references}

\clearpage

\appendix
\onecolumn

\section{Full Single-Part Results}\label{ablation}
{\tiny
\begin{table}[!htbp]
  \centering
  \setlength\tabcolsep{3pt}
  \begin{tabular}{rrllrr}
    \toprule
    \cmidrule{1-6}
    \# & Params & Rhythm Model & Notes Model & $\text{Loss}_t$ & $\text{Loss}_n$ \\
    \midrule
    1 & 112 & $\hat{\textbf{r}}_{k,0} = \textbf{bias}_0$ & $\hat{\textbf{r}}_{k,1,n} = \textbf{bias}_{1,n}$ & 2.92 & 7.15 \\
    2 & 21k & $\hat{\textbf{r}}_{k,0} = \textbf{lin}(\textbf{r}_{(1)})$ & $\hat{\textbf{r}}_{k,1,n} = \textbf{lin}_n(\textbf{r}_{(1)},\textbf{r}_+) $ & 2.00 & 6.05 \\
    3 & 9k & $\hat{\textbf{r}}_{k,0} = \textbf{lin}(\textbf{r}_{(1)})$ & $\hat{\textbf{r}}_{k,1,n} = \textbf{lin}(\textbf{r}_{(1)},\textbf{r}_+) $ & 2.00 & 4.29  \\
    4 & 11k & $\hat{\textbf{r}}_{k,0} = \textbf{lin}(\textbf{r}_{(1)},\ell)$ & $\hat{\textbf{r}}_{k,1,n} = \textbf{lin}(\textbf{r}_{(1)},\textbf{r}_+,\textbf{1}_n) $ & 1.83 & 4.29  \\
    5 & 149k & $\hat{\textbf{r}}_{k,0} = \textbf{lin}\circ \textbf{fc}(\textbf{r}_{(1)})$ & $\hat{\textbf{r}}_{k,1,n}  = \textbf{lin}_n \circ \textbf{fc}(\textbf{r}_{(1)},\textbf{r}_+) $ & 1.99 & 3.93  \\
    6 & 135k & $\hat{\textbf{r}}_{k,0} = \textbf{lin}\circ \textbf{fc}(\textbf{r}_{(1)})$ & $\hat{\textbf{r}}_{k,1,n} = \textbf{lin} \circ \textbf{fc}(\textbf{r}_{(1)},\textbf{r}_+) $ & 1.99 & 4.07  \\
    7 & 172k & $\hat{\textbf{r}}_{k,0} = \textbf{lin}\circ\textbf{fc}(\textbf{r}_{(1)},\ell)$ & $\hat{\textbf{r}}_{k,1,n}  = \textbf{lin} \circ \textbf{fc}(\textbf{r}_{(1)},\textbf{r}_+,\textbf{1}_n) $ & 1.80 & 3.90  \\
    \midrule
    \midrule
    8 & 72k & $\hat{\textbf{r}}_{k,0} = \textbf{lin}(\textbf{r}_{(5)})$ & $\hat{\textbf{r}}_{k,1,n} = \textbf{lin}_n(\textbf{r}_{(5)},\textbf{r}_+) $ & 1.86 & 6.05 \\
    9 & 36k & $\hat{\textbf{r}}_{k,0} = \textbf{lin}(\textbf{r}_{(5)})$ & $\hat{\textbf{r}}_{k,1,n} = \textbf{lin}(\textbf{r}_{(5)},\textbf{r}_+) $ & 1.86 & 3.91  \\
   10 & 38k & $\hat{\textbf{r}}_{k,0} = \textbf{lin}(\textbf{r}_{(5)},\ell)$ & $\hat{\textbf{r}}_{k,1,n} = \textbf{lin}(\textbf{r}_{(5)},\textbf{r}_+,\textbf{1}_n) $ & 1.73 & 3.91  \\
    11 & 418k & $\hat{\textbf{r}}_{k,0} = \textbf{lin}\circ \textbf{fc}(\textbf{r}_{(5)})$ & $\hat{\textbf{r}}_{k,1,n} = \textbf{lin}_n \circ \textbf{fc}(\textbf{r}_{(5)},\textbf{r}_+) $ & 1.64 & 3.26  \\
    12 & 497k & $\hat{\textbf{r}}_{k,0} = \textbf{lin}\circ \textbf{fc}(\textbf{r}_{(5)})$ & $\hat{\textbf{r}}_{k,1,n} = \textbf{lin} \circ \textbf{fc}(\textbf{r}_{(5)},\textbf{r}_+) $ & 1.64 & 3.16  \\
    13 & 535k & $\hat{\textbf{r}}_{k,0} = \textbf{lin}\circ\textbf{fc}(\textbf{r}_{(5)},\ell)$ & $\hat{\textbf{r}}_{k,1,n} = \textbf{lin} \circ \textbf{fc}(\textbf{r}_{(5)},\textbf{r}_+,\textbf{1}_n) $ & 1.59 & 3.10  \\
    14 & 228k & $\hat{\textbf{r}}_{k,0} = \textbf{lin}\circ\textbf{fc}(\textbf{f}(\textbf{r}_{(5)}),\ell)$ & $\hat{\textbf{r}}_{k,1,n}= \textbf{lin} \circ \textbf{fc}(\textbf{c}(\textbf{r}_{(5)}),\textbf{r}_+,\textbf{1}_n) $ & 1.58 & 3.05  \\
    \midrule
    \midrule
    15 & 134k & $\hat{\textbf{r}}_{k,0} = \textbf{lin}(\textbf{r}_{(10)})$ & $\hat{\textbf{r}}_{k,1,n} = \textbf{lin}(\textbf{r}_{(10)},\textbf{r}_+) $ & 1.83 & 6.05 \\
    16 & 71k & $\hat{\textbf{r}}_{k,0} = \textbf{lin}(\textbf{r}_{(10)},\ell)$ & $\hat{\textbf{r}}_{k,1,n} = \textbf{lin}(\textbf{r}_{(10)},\textbf{r}_+,\textbf{1}_n) $ & 1.71 & 3.83 \\
    17 & 372k & $\hat{\textbf{r}}_{k,0} = \textbf{lin}\circ\textbf{fc}(\textbf{f}(\textbf{r}_{(10)}),\ell)$ & $\hat{\textbf{r}}_{k,1,n} = \textbf{lin} \circ \textbf{fc}(\textbf{c}(\textbf{r}_{(10)}),\textbf{r}_+,\textbf{1}_n) $ & 1.55 & 3.00  \\
    18 & 250k & $\hat{\textbf{r}}_{k,0} = \textbf{lin}\circ \textbf{conv}_5(\textbf{f}(\textbf{r}_{(10)}),\ell)$ & $\hat{\textbf{r}}_{k,1,n} = \textbf{lin}\circ\textbf{conv}_5(\textbf{c}(\textbf{r}_{(10)}),\textbf{r}_+,\textbf{1}_n) $ & 1.55 & 3.01  \\
    19 & 769k & $\hat{\textbf{r}}_{k,0} = \textbf{lin}\circ \textbf{conv}_3\circ \textbf{conv}_5(\textbf{f}(\textbf{r}_{(10)}),\ell)$ & $\hat{\textbf{r}}_{k,1,n} = \textbf{lin}\circ\textbf{conv}_3 \circ \hspace{1mm} \textbf{conv}_5(\textbf{c}(\textbf{r}_{(10)}),\textbf{r}_+,\textbf{1}_n) $ & 1.50 & 2.92  \\
     20 & 342k & $\hat{\textbf{r}}_{k,0} = \textbf{lin}\circ \textbf{rnn}(\textbf{r}_{(10)},\ell)$ & $\hat{\textbf{r}}_{k,1,n} = \textbf{lin}\circ\textbf{rnn}(\textbf{r}_{(10)},\textbf{r}_+,\textbf{1}_n))$ & 1.48 & 2.89 \\
     21 & 283k & $\hat{\textbf{r}}_{k,0} = \textbf{lin}\circ \textbf{rnn}(\textbf{f}(\textbf{r}_{(10)}),\ell)$ & $\hat{\textbf{r}}_{k,1,n} = \textbf{lin}\circ\textbf{rnn}(\textbf{c}(\textbf{r}_{(10)}),\textbf{r}_+,\textbf{1}_n))$ & 1.48 & 2.88 \\
    \midrule
    \midrule
     22 & 301k & $\hat{\textbf{r}}_{k,0} = \textbf{lin}\circ \textbf{rnn}(\textbf{f}(\tilde{\textbf{r}}_{(10)}),\ell)$ & $\hat{\textbf{r}}_{k,1,n} = \textbf{lin}\circ\textbf{rnn}(\textbf{c}(\tilde{\textbf{r}}_{(10)}),\textbf{r}_+,\textbf{1}_n))$ & 1.59 & 2.93 \\
    \bottomrule
  \end{tabular}
  \caption{Single-part results. Loss is the cross-entropy described in Section 3.1. Loss$_t$ and Loss$_n$ are decompositions of the loss described in Section 3.2. For succinctness, define $\textbf{r}_{(m)} \equiv \textbf{r}_{k-m:k}$ (a truncated history of length $k$) and $\textbf{r}_+ \equiv \textbf{r}_{k,0} \oplus \textbf{r}_{k,1,1:n}$ (the current frame, masked above pitch $n$). For definition of \textbf{r} see Section 4, Sequential part factorization. $\textbf{lin}_n$ indicates a log-linear classifier (sigmoid for $\hat y_n$ and softmax for $\hat y_t$) and \textbf{lin} indicates the relative pitch log-linear classifier and inputs $\textbf{1}_n$ indicate pitch-class features (Section 5.2, Relative pitch). The inputs $\ell$ indicate location features (Section 5.2, Autoregressive modeling). \textbf{fc} indicates a fully connected layer. \textbf{f} and \textbf{c} indicates pitch embeddings (Section 5.2, Pitch embeddings). $\textbf{conv}_k$ indicates 1d convolution of width $k$. \textbf{rnn} indicates a recurrent layer. All hidden layers are parameterized with 300 nodes. Models were regularized with early stopping when necessary. The subscript $k$ on the history tensor $x_k$ indicates the number of frames of history used in each experiment (either 1, 5, or 10 frames). $\tilde{\textbf{r}}_{(m)}$ is a modified history discussed in Section 5.1. } 
 \label{homophonicresults}
\end{table}
}
\section{Piano Music}\label{pianomusic}

\begin{figure}
  \centering
  \includegraphics[width=0.9\textwidth]{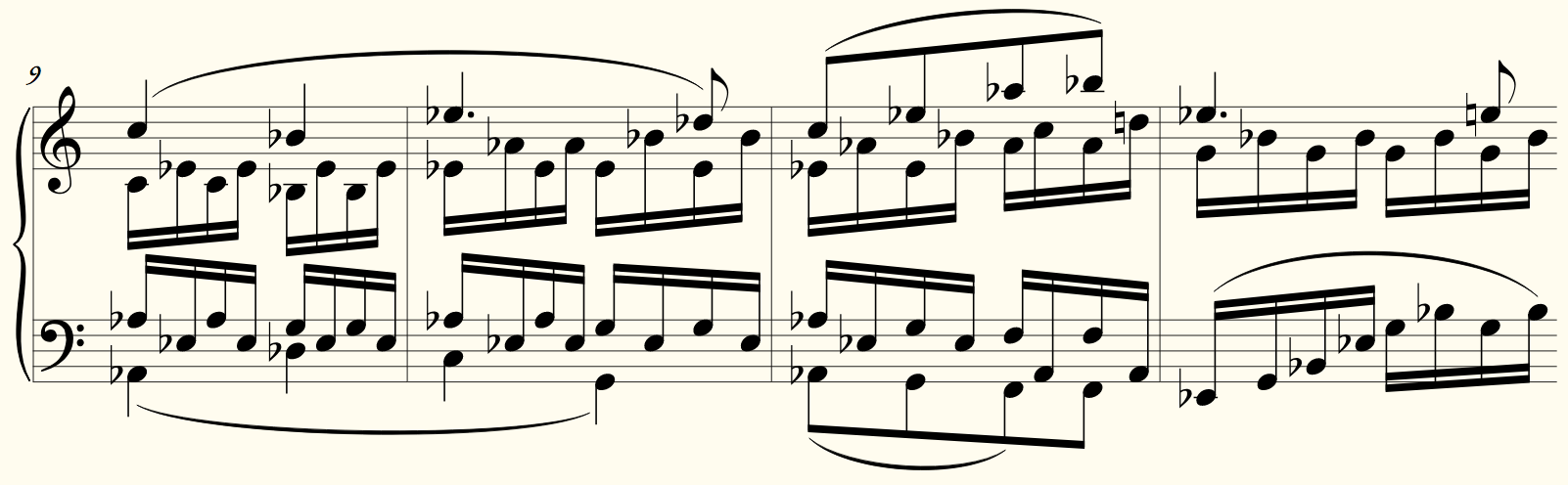}
    \caption{Beethoven's piano sonata number 8 (Pathetique) movement 2, from measure 9, rendered by the Verovio Humdrum Viewer. Although visually rendered on two staves, this sonata consists of four parts: a high sequence of quarter and eighth notes, two middle sequences of sixteenth notes, and a low sequence of quarter notes.}\label{moonlight}
\end{figure}

For some piano music, it is necessary to draw a distinction between an instrument and a part. Consider the piano score given in Figure \ref{moonlight}. This single piano part is more comparable to a complete score than the individual parts of, for example, a string quartet (compare the piano score in Figure \ref{moonlight} to the quartet score in Figure 1 in the main text). Indeed, an educated musician would read this score in four distinct parts: a high sequence of quarter and eighth notes, two middle sequences of sixteenth notes, and a low sequence of quarter notes. In measure 12, the lowest two parts combine into a single bass line of sixteenth notes.

These part divisions are indicated in score through a combination of beams, slurs, and other visual queues. We do not model these visual indicators; instead we rely on part annotations provided by the KernScores dataset. The provision of these annotations is a strong point in favor of the KernScores dataset's Humdrum format; although in principle formats like MIDI can encode this information, in practice they typically collect all notes for a single instrument into a single track, or possibly two tracks (for the treble and bass staves, as seen in the figure) in the case of piano music.

In extremely rare cases, this distinction between instrument and part must also be made for stringed instruments; a notable example is Beethoven's string quartet number 14, in the fourth movement in measures 165 and 173, where the four instruments each separate into two distinct parts creating brief moments of 8-part harmony. The physical constraints of stringed instruments discourage more widespread use of these polyphonies. For vocal music, of course, physical constraints prevent intra-instrument polyphony entirely.

As Figure \ref{moonlight} illustrates, these more abstract parts can weave in and out of existence. Two parts can merge with each other; a single part can split in two; new parts can emerge spontaneously. The KernScores data provides annotations that describe this behavior. We can represent these dynamics of parts as a $P \times P$ flow matrix at each time step ($P$ is an upper bound on the number of parts; for the KernScores corpus used in this work, we take $P = 6$) that describes where each part moves in the next step. At most time steps, this flow matrix is the identity matrix.

The state-based models discussed in this paper can easily be adjusted to accommodate these flows. If two parts merge, sum their states; if a part splits in two, duplicate its state. These operations amount to hitting the vector of state estimates for the parts with the flow matrix at each time step. However, we do not currently model the flow matrix. Because the flow matrix for piano music contains some (small) amount of entropy, we therefore exclude piano music from the results reported in Table 4. We do however include the piano music in training. 

\section{Piano-Roll Representations of Scores}\label{factormath}

\begin{figure}[h]
  \centering
  \includegraphics[width=0.36\textwidth]{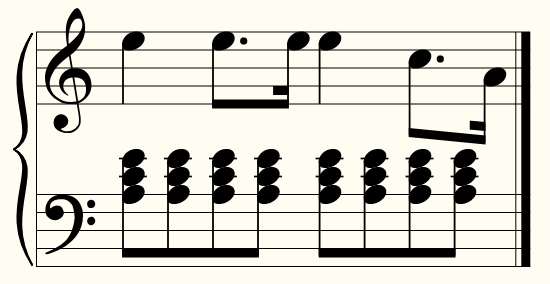}
    \caption{Mozart's piano sonata number 8 in A minor, movement 1, from measure 1.}
\end{figure}

In Section 3.1 we defined a score as a $T \times P \times 2N$ binary tensor, where at each time $t \in T$ in each part $p \in P$, we have two values $\textbf{x}_{t,p,n}$ and $\textbf{x}_{t,p,N+n}$ to indicate whether note $n\in N$ is present and whether $n$ begins respectively. While classical piano-roll representations omit the second onset bit, both bits are necessary to faithfully represent a musical score. Consider, for example, Figure C.1. Two scores are demonstrated in Figure C.2 that have identical piano-roll encodings if only a single bit is used to indicate the presence of a note. Many other scores also alias to this same piano-roll encoding. The addition of an onset bit delineates the boundaries between multiple notes of the same pitch, thus resolving this ambiguity.

\begin{figure}[h]
\begin{minipage}{0.5\textwidth}
\center
\includegraphics[width=0.6\linewidth,scale=1.0]{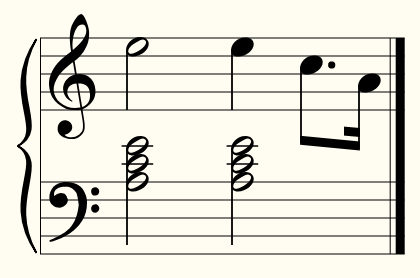}
\end{minipage}
\begin{minipage}{0.5\textwidth}
\center
\includegraphics[width=0.6\linewidth,scale=1.0]{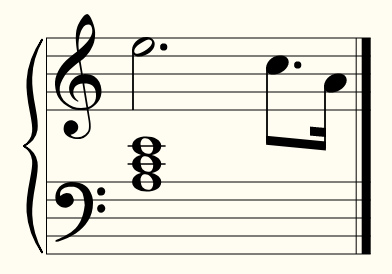}
\end{minipage}
\caption{Two scores with the same piano-roll representation as the score in Figure B.2. The popular dataset introduced by Boulanger-Lewandowski et al. (2012) uses this single-bit representation. A second bit is used in some more recent work, for example Liang et al. (2017) in which they are referred to as ``Tie'' bits).}
\label{confuse}
\end{figure}

Another pitfall of piano-roll representations is the choice of discretization. In Section 3.1, we defined a continuous-time process with a real-valued index $t$. To use a piano-roll for factorization or featurization, a finite resolution must be chosen. We argued in Section 3.2 that this discretization $\Delta$ is information-preserving, so long as $\Delta$ is chosen to be the resolution of the score process or finer. The consequences of choosing a discretization that is too coarse is illustrated by Figure C.3.

\begin{figure}
  \centering
  \includegraphics[width=0.36\textwidth]{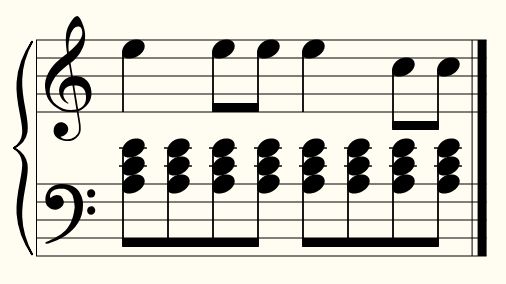}
    \caption{A corruption of the score from Figure C.1, discretized at eighth-note resolution.}\label{coarse}
\end{figure}

\section{Run-Length Factorization}

Training and sampling from a model over a discrete factorization of scores at the process resolution $\Delta$ can be expensive, prompting some earlier works to discretize at a coarser resolution (as discussed in the previous section). One approach to preserve fine rhythmic structure (e.g. triplets and thirty-second notes) without committing to a fine discretization is to factor a score into run-lengths. To this end, we define a run-length encoded score $\textbf{x} \in \left(\mathbb{N} \oplus \{0,1\}^{P \times 2N}\right)^T$ where, at each time index $t \in \{1,\dots,T\}$, we set
\begin{align*}
& \textbf{x}_{t,0} = \textbf{1}_{d_t}, & & \text{where $d_t$ is the duration of the event at time index $t$},\\
& \textbf{x}_{t,1,p,n} = 1 & & \text{iff note $n$ is on at time $t$ in part $p$},\\
& \textbf{x}_{t,1,p,2n} = 1 & & \text{iff note $n$ begins at time $t$ in part $p$}.
\end{align*}
The sequence $\textbf{x}_t$ is non-linear in the index $t$: entry $\textbf{x}_{t+1}$ occurs $d_t$ beats after entry $\textbf{x}_t$, in contrast to the raster where $\textbf{x}_{t+1}$ always occurs a constant interval $\Delta$ after $\textbf{x}_t$.

We can then factor the distribution over scores into conditional distributions over binary note values and natural-number duration values:
\begin{equation*}
p(S) = \prod_{t=1}^T p(\textbf{x}_t|\textbf{x}_{1:t}) = \prod_{t=1}^T p(\textbf{x}_{t,0}|\textbf{x}_{1:t}) \prod_{p=1}^P\prod_{n=1}^{2N} p(\textbf{x}_{t,p,n}|\textbf{x}_{1:t},\textbf{x}_{t,0},\textbf{x}_{t,1,1:p},\textbf{x}_{t,1,p,1:n}).
\end{equation*}
Because music typically doesn't evolve at the finest possible resolution $\Delta$, we save a substantial amount of computation by predicting run-lengths $\textbf{x}_{t,0} \in \mathbb{N}$ rather than re-iterating the predictions $\textbf{x}_{t,p,n}$ at successive time steps.

\begin{figure}
  \centering
  \includegraphics[width=0.4\textwidth]{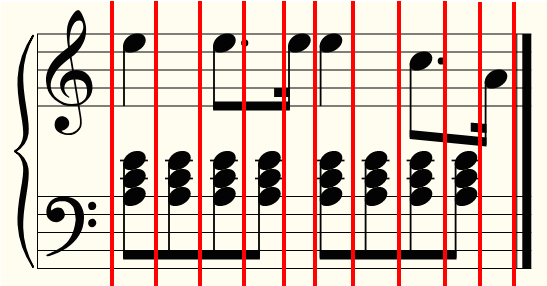}
    \caption{The Mozart from Figure C.1, with red lines that indicate the boundaries of events under a run-length factorization of the score. Notes in the treble staff are chopped up into eight-note runs, so instead of predicting note durations (quarter, dotted-eighth, sixteenth, etc.) we instead predict fragments of notes (eighth, continue eighth, continue eighth, etc.).}\label{mozartevents}
\end{figure}

One criticism of the run-length factorization is that, when notes of different durations overlap in a score, the longer notes are chopped up along the boundaries of the short notes as illustrated in Figure \ref{mozartevents}. Rather than predicting musically meaningful quantities like note values (quarter, eighth, dotted-eighth, etc.) instead we predict run-length chunks.

\clearpage

\section{User Study Details}\label{userstudy}

To understand our model qualitatively, we asked 20 study participants to evaluate compositions produced by one of our best models: experiment 6 from Table 4. Each user was asked to listen to 5 sets of 10 audio clips, synthesized from scores ranging from 10  to 50 frames of composition (a frame is a run-length as defined by the representation discussed in Section 5.1). Every user was presented with their own set of audio clips, randomly sampled from either the training set or the model. Users were given the following prompt before beginning the study:

\begin{quote}
This is a musical Turing test. You will be presented with a selection of audio clips, beginning with short clips and progressing to longer clips. For each audio clip, you will be asked whether you believe the clip was composed by a human or a computer. Half the clips you will be presented with belong in each category. This data contains many famous classical compositions, ranging from the Renaissance to early 20th century. If you specifically recognize a piece, please let me know. Finally, all recordings you hear--both human and artificial--are performed at a tempo of 120bpm.
\end{quote}

Additionally, we asked users two questions about their background:
\begin{itemize}
\item Do you self-identify as musically educated? (8 responded `yes')
\item Do you self-identify as educated in machine learning? (13 responded `yes')
\end{itemize}

Table 2 summarizes results of our listening study, including conditional results for the educated subgroups.

{\tiny
\begin{table}[h]
  \centering
  \begin{tabular}{lrrrrr}
    \toprule
    \cmidrule{1-6}
     Frames & 10 & 20 & 30 & 40 & 50 \\
     \midrule
     All & 5.3 & 5.7 & 6.6 & 6.7 & 6.8 \\
     Music & 4.9 & 6.0 & 6.4 & 6.9 & 7.0 \\
     ML & 4.8 & 5.5 & 6.2 & 6.7 & 6.8 \\
    \bottomrule
  \end{tabular}
  \caption{Qualitative evaluation of the 10-frame hierarchical model: Experiment 6 in Table 4. Twenty participant were asked to judge 50 audio clips each of varying length. The scores indicate participants' average correct discriminations out of 10 (5.0 would indicate random guessing; 10.0 would indicate perfect discrimination). The categories indicate breakdowns for listeners who identified as educated in music or educated in machine learning. }
\end{table}}

We asked users to identify pieces if they specifically recognized them, because we were concerned that this knowledge of the classical music canon could confound the question of musical plausibility of our model's samples. In the end, only one user positively identified a piece in our study. This may be explained because our models do not predict tempo. Therefore, to make fair comparisons between human compositions and model outputs, we synthesized all scores at a tempo of 120bpm. This may serve to obscure recognizable pieces. However, it also makes the task less informative because all audio clips in the listening test sound less like ``real music.'' Participants were informed of this fact, but it is not clear how effectively they could use this knowledge.
\section{Evaluation Metric Details}\label{appeval}

We demonstrate here the claim from the text that any refinement of the $\Delta$-interval constant discretization of the support of the distribution $p$ over scores yields no further contributions to the cross entropy. Formally, if $\mathcal{P} = (0,\Delta,2\Delta,\dots,T)$ is the discrete partition of the interval $[0,T]$ and $\mathcal{R}$ is any refinement of this partition (i.e. a partition of $[0,T]$ such that contains the points of $\mathcal{P}$) then the following proposition holds.

\begin{prop}\label{prop:res}
For any refinement $\mathcal{R}$ of $\mathcal{P}$, $H_\mathcal{P}(p||q) = H_\mathcal{R}(p||q)$.
\end{prop}
\begin{proof}
Let $K'$ denote the size of $\mathcal{R}$. By definition of relative entropy of $p$ restricted to the partition $\mathcal{R}$,
\begin{equation*}
H_\mathcal{R}(p||q) = \underset{\textbf{x} \sim p}{\mathbb{E}} \log q(\textbf{x}_{\mathcal{R}_1},\textbf{x}_{\mathcal{R}_1},\dots,\textbf{x}_{\mathcal{R}_{K'}}).
\end{equation*}
And applying the chain rule for conditional probabilities, 
\begin{equation*}
 H_\mathcal{R}(p||q) = \sum_{k=1}^{K'} \underset{\textbf{x} \sim p}{\mathbb{E}}\log q(\textbf{x}_{\mathcal{R}_k}|\textbf{x}_{\mathcal{R}_1},\dots,\textbf{x}_{\mathcal{R}_{k-1}}).
 \end{equation*}
Consider terms $q(\textbf{x}_{\mathcal{R}_k}|\textbf{x}_{\mathcal{R}_1},\dots,\textbf{x}_{\mathcal{R}_{k-1}})$ where $\mathcal{R}_k \notin \mathcal{P}$. There exists some $n$ such that $n\Delta < \mathcal{R}_k < (n+1)\Delta$. We must have $\textbf{x}_{\mathcal{R}_k} = \textbf{x}_{n\Delta}$ because by definition of $\Delta$, all change-points in \textbf{x} occur at integer multiples of $\Delta$. Because $\mathcal{R}$ is a refinement of $\mathcal{P}$ and $n\Delta \in \mathcal{P}$, it follows that $n\Delta \in \mathcal{R}$. Furthermore, $n\Delta < \mathcal{R}_k$ and therefore $n\Delta \in (\mathcal{R}_0,\dots,\mathcal{R}_{k-1})$. We conclude that
\begin{equation*} \underset{\textbf{x} \sim p}{\mathbb{E}} \log q(\textbf{x}_{\mathcal{R}_k}|\textbf{x}_{\mathcal{R}_1},\dots,\textbf{x}_{\mathcal{R}_{k-1}}) = \underset{\textbf{x} \sim p}{\mathbb{E}}\log q(\textbf{x}_{\mathcal{R}_k}|\textbf{x}_{n\Delta},\dots) = 0.
\end{equation*}
In words: conditioned on $\textbf{x}_{n\Delta}$, $\textbf{x}_{\mathcal{R}_k}$ is known and its relative entropy vanishes. Dropping all such terms $k$ with $\mathcal{R}_k \notin \mathcal{P}$ we see that
\begin{equation*}
 H_\mathcal{R}(p||q) = \sum_{k \text{ : } \mathcal{R}_k \in \mathcal{P}} \underset{\textbf{x} \sim p} {\mathbb{E}}\log q(\textbf{x}_{\mathcal{R}_k}|\textbf{x}_{\mathcal{R}_0},\dots,\textbf{x}_{\mathcal{R}_{k-1}})
 \end{equation*}
 \begin{equation*} 
 = \sum_{k=1}^{T/\Delta} \underset{\textbf{x} \sim p} {\mathbb{E}}\log q(\textbf{x}_{k\Delta} |\textbf{x}_0,\dots,\textbf{x}_{(k-1)\Delta}) = \underset{\textbf{x} \sim p}{\mathbb{E}} \log q(\textbf{x}_0,\textbf{x}_\Delta,\dots,\textbf{x}_T) =  -T\Delta H(p||q). \qedhere
\end{equation*}
\end{proof}

The point of this calculation is that, beyond some level of refinement, further increasing the resolution of the score process yields no further contributions to the entropy; the intermediate frames are completely determined by their neighbors. It may be illuminating to draw a contrast here with a truly continuous process such as Brownian motion, for which further refinement of the sampling partition continues to yield new details of the process at any resolution.

% For non bibtex users:
%\begin{thebibliography}{citations}
%
%\bibitem {Author:00}
%E. Author.
%``The Title of the Conference Paper,''
%{\it Proceedings of the International Symposium
%on Music Information Retrieval}, pp.~000--111, 2000.
%
%\bibitem{Someone:10}
%A. Someone, B. Someone, and C. Someone.
%``The Title of the Journal Paper,''
%{\it Journal of New Music Research},
%Vol.~A, No.~B, pp.~111--222, 2010.
%
%\bibitem{Someone:04} X. Someone and Y. Someone. {\it Title of the Book},
%    Editorial Acme, Porto, 2012.
%
%\end{thebibliography}

\end{document}